\documentstyle[aps,epsf,prc]{revtex}

\begin{document}
\title{The reaction $\Delta+N\to N+N+\phi$ in ion-ion collisions}
\author{Michail P. Rekalo \footnote{ Permanent address:
\it National Science Center KFTI, 310108 Kharkov, Ukraine}
}
\address{Middle East Technical University, 
Physics Department, Ankara 06531, Turkey}
\author{Egle Tomasi-Gustafsson}
\address{\it DAPNIA/SPhN, CEA/Saclay, 91191 Gif-sur-Yvette Cedex, 
France}
\date{\today}

\maketitle
\begin{abstract}
We study the threshold $\phi$-meson production in the process $\Delta+N\to N+N+\phi$, which appears as a possible important mechanism in high energy nuclei-nuclei collisions. The isotopic invariance of the strong interaction and the selection rules due to P-parity and total angular momentum result in a general and model independent parametrization of the spin structure of the matrix element in terms of three partial amplitudes. In the framework of one-pion exchange model these amplitudes can be derived in terms of the two threshold  partial amplitudes for the process $\pi+N\to N+\phi$. We predict the ratio of cross sections for $\phi-$meson production in $pp$- and $\Delta N$-collisions and the polarization properties of the $\phi$-meson, in $\Delta+N\to N+N+\phi$, as a function of a single parameter, which characterizes the relative role of transversal and longitudinal $\phi$-meson polarizations in the process $\pi+N\to N+\phi$.

\end{abstract}
\section{Introduction}
Experimental \cite{Ba01} and theoretical \cite{Si97,Si00,Na98,Na99,Ti00} studies devoted to $\phi$-meson production in nucleon-nucleon collisions have known a growing interest in recent years.  In particular efforts have been focussed on the understanding of the reaction mechanism for $N+N\to N+N+\phi$, which is one of the simplest processes of vector meson production in $NN$-collisions. The standard approach, based on one-boson exchanges in $t-$channel (the peripherical picture) is not unique, in particular in the threshold region. A picture based on central $NN$-collisions, with $s-$channel excitation of six-quark bags \cite{Fa01}, can also be applied to vector meson production in $NN$-collisions \cite{Ta00}.

It has been pointed out that the process $N+N\to N+N+\phi$, and in particular the study of polarization phenomena in $p+p\to p+p+\phi$ can bring information  on the hidden strangeness degrees of freedom in the nucleon \cite{El95}. This problem is related with the possible violation of the OZI-rule \cite{OZI} for various processes of hadronic $\omega$ and $\phi$-production. In the framework of specific models \cite{Na98,Na99,Ti00} in the near threshold region, it would be possible to determine or to constrain the coupling constants of the $NN\phi$-interaction, presently unknown.

A suggestion was made \cite{Sh85}, that a strong enhancement of the $\phi/\omega$ ratio in heavy ion collisions, in comparison with the value measured in $p+p$-collisions, could be interpreted as a signature of quark-gluon plasma formation. Search for such signal has been done in the CERN experiment NA38 \cite{Ba91}.

These examples show that the comprehension of the elementary process $N+N\to N+N+\phi$ is important for nucleon-nucleon collisions and nucleon-nucleus collisions, as well. However, the reaction $\Delta+N\to N+N+\phi$, where the nucleon participates in its first excited state, may also contribute to $\phi$-production in nuclei, in non negligible way. At bombarding energies of about 1 GeV/A, a particular state of nuclear matter can be formed \cite{Me93}, where about 1/3 of baryons are excited to the $\Delta$-resonance state. It follows that the $\phi$-meson in $p+A$- or $A+A$-collisions can be produced in non negligeable amount, through the process $\Delta+N\to N+N+\phi$. The importance of $\Delta N$-reactions for the $\eta$-meson production has been considered earlier \cite{Pe96}.

The aim of this paper is to study $\phi-$production in the process: $\Delta +N\to N+N+\phi$, in particular in the near threshold region. The threshold region is interesting from an experimental point of view and the theoretical approach can be essentially simplified. The presence of the $\Delta$-isobar, with 3/2 spin, makes the situation more complicated than in case of the nucleon, but also for  the  $\Delta +N$ interaction it is possible to develop a formalism, similar to $\phi$-meson production in $NN$ interaction \cite{Re97a}. The symmetry properties of the strong interaction, such as the Pauli principle for the nucleons, the P-invariance, the isotopic invariance and the selection rules for the total angular momentum, allow to define the spin structure of the matrix element for $\Delta+N\to N+N+\phi$ in terms of a small number of partial amplitudes. This approach is especially adapted to the near-threshold region, where all final particles are produced in S-state. The existing experimental data about $\phi-$ and $\omega$-production in $NN-$collisions \cite{Bo68,Co67,Ca68,Co70,Ye70,Al68,Hi99,Ab01} show that this region is quite wide. There is no particular reason to expect a different behavior in case of $\phi$-production in $\Delta +N$-collisions. On the contrary, taking into account the hidden strangeness of the $\phi$-meson, due to the relatively small $strange$ radius of the nucleon \cite{Re97b}, the $S-$wave region for $\phi$-production should be larger than  for $\omega$-production. A similar expectation has been also confirmed by experimental data on associative strange particle production in $NN-$collisions, in the near threshold region 
\cite{Ba98,Bi98,Se98,Ba99}.

This paper is organized as follows. In the first chapter we establish the spin structure of the threshold matrix for the process: $\Delta+N\to N+N+\phi$. Due to the isotopic invariance of the strong interaction, this spin structure applies to all possible combinations of the electric charge of baryons:
$\Delta^{++}+n\to p+p+\phi$, $\Delta^+ +p\to p+p+\phi$, $\Delta^+ +n\to n+p+\phi$, $\Delta^0 +p\to n+p+\phi$, 
$\Delta^0 +n\to n+n+\phi$, and  $\Delta^- +p\to n+n+\phi$, because all these processes, in the $s$-channel, have a single value of isotopic spin, $I=1$. The isotopic invariance is also very important with respect to the selection rules related to the P-parity and the total angular momentum. In Chapter II, using this parametrization, we show that any observable for the process $\Delta+N\to N+N+\phi$ can be calculated in terms of three independent partial amplitudes, which characterize this process in the threshold region. From all possible polarization observables, only the density matrix of the $\phi$-meson has a physical meaning, as this process can in principle occur only inside nuclei.

After this model independent analysis, we consider the one-boson model, with $\pi$-exchange, taking into account the part of this mechanism related to the subprocess $\pi+N\to N+\phi$. This allows us to compare the present calculation with calculations on $ N +N\to N+N+\phi$-processes, without addition of new parameters, in the framework of the same model. The results are summarized in the Conclusions.
\section{Spin structure of threshold matrix element}
Considering the $\Delta$-isobar as a particle with spin 3/2 and positive P-parity, it is possible to find the spin structure of the process
 $\Delta+N\to N+N+\phi$ in threshold regime. The threshold region can be rigorously defined in the center of mass system (CMS) of the considered reaction in terms of orbital angular momenta. If $\ell_1$ is the relative orbital momentum of the final nucleons and $\ell_2$ the orbital momentum of the $\phi$-meson, the condition of the $S-$wave production of any pair of particles for the $NN\phi$ final state can be written as $\ell_1=\ell_2=0$.
 
As mentioned in the Introduction, the threshold region may be quite wide for the $\phi$-meson, due to the fact that the region of the nucleon responsible for strange particle production is small. Assuming  the strange radius $R_S\simeq 1/m_K$, where $m_K$ is the kaon mass, semi-classical considerations suggest that for $q\le m_K$ (q is the $\phi$-meson momentum in CMS) the $S-$approximation is justified. Such consideration was applied in the phenomenological analysis \cite{Ba01} of $\phi$-meson production in $NN$-collisions.

The spin structure of the threshold matrix element for any process $\Delta+N\to N+N+\phi$ can be established in a general form, from the selection rules in isotopic spin, $P$-parity and the total angular momentum. From the isotopic invariance of the strong interaction, one can find that any pair of final nucleons in all reactions $\Delta+N\to N+N+\phi$ (for any charge combination, $pp$, $nn$ or $np$) has to be in singlet state, following the generalized Pauli principle. Due to the conservation of the total isotopic spin, $I$, the final $NN-$system must have $I=1$. Therefore, the total angular momentum ${\cal J}$ and the $P$-parity of the entrance channel can take only the value ${\cal J}^P=1^-$. Taking into account the conservation of ${\cal J}$ and $P$, one can find the following allowed partial transitions for the process  $\Delta+N\to N+N+\phi$:
\begin{eqnarray}
S_i=1,~\ell=1~&\to& ~{\cal J}^{ P}=1^-, \nonumber \\
S_i=2,~\ell=1~&\to& ~{\cal J}^{ P}=1^-,
\label{eq:tran}\\
S_i=2,~\ell=3~&\to& ~{\cal J}^{ P}=1^-, \nonumber
\end{eqnarray}
where $S_i$  is the total spin of the initial $\Delta+N$-system and 
$\ell$ is the angular orbital momentum of this system.

The threshold matrix element, in the CMS of the considered reaction,  can be parametrized in the following general form:
\begin{eqnarray}
&{\cal M}=&if_1(\tilde{\chi}_1~\sigma_y ~\vec\Delta\cdot\hat{\vec k }
\times\vec{U^*} )~
(\chi^{\dagger}_3\sigma_y \tilde{\chi}^{\dagger}_2 )\nonumber\\
&&+f_2(\tilde{\chi}_1~\sigma_y \vec\sigma \cdot\vec{U^*}\vec\Delta \cdot\hat{\vec k } )~
(\chi^{\dagger}_3\sigma_y \tilde{\chi}^{\dagger}_2 )\label{eq:mel}\\
&&+f_3(\tilde{\chi}_1~\sigma_y \vec\sigma \cdot\hat{\vec k }\vec\Delta\cdot\hat{\vec k } )~
(\chi^{\dagger}_3\sigma_y \tilde{\chi}^{\dagger}_2 )\hat{\vec k }
\cdot\vec{U^*},\nonumber
\end{eqnarray}
where $\chi_1$ ($\chi_2$ and $\chi_3$) is the
two-component spinor of the initial (final) nucleon;  
$\vec\Delta$ is a particular two-component spinor (each component is a vector) for the description of the polarization properties of the $\Delta$-isobar (with spin 3/2), so that $\vec\sigma\cdot\vec\Delta=0$,
$\vec  U$ is the three-vector of the $V$-meson polarization, and $\hat{\vec k}$ is
the unit vector along the 3-momentum of the $\Delta$; $f_1-f_3$ are the threshold partial amplitudes describing the allowed transitions (\ref{eq:tran}).

The presence of the Pauli matrix $\sigma_y$ in the parametrization (\ref{eq:mel}) insures the correct transformation properties of the corresponding two-component spinor products, relative to rotation. 

Taking into account the conservation of isospin, one obtains:
$$ -{\cal M}(\Delta^+p\to pp\phi)=
\displaystyle\frac{1}{\sqrt{3}} {\cal M}(\Delta^{++}n\to pp\phi)=
{\cal M}(\Delta^{+}n\to np\phi)=
-{\cal M}(\Delta^0 p\to np\phi),$$
i.e. the same set of three amplitudes $f_1-f_3$ describes all these processes and therefore the polarization phenomena have to be identical for all reactions $\Delta+N\to N+N+\phi$.

After summing over the polarizations of the final particles and averaging over the polarizations of the colliding baryons, one can find the following formula for the differential cross section of the considered process, in terms of the partial amplitudes $f_1-f_3$:
\begin{equation}
\displaystyle\frac{d\sigma}{d\omega}=\displaystyle\frac{1}{6}{\cal N}
\left ( |f_1-f_2|^2+3|f_1+f_2|^2+2|f_2+f_3|^2\right ),
\label{eq:sig}
\end{equation}
$${\cal N}=2m(E_N+m)(E_{\Delta}+m),$$
where  $d\omega$ is the element of phase-space for the three-particle final state. $E_N$ and $E_{\Delta}$ are the energies of the colliding $N$ and $\Delta$ at threshold, in CMS:
$$E_N=\displaystyle\frac{W^2_{th}+m^2-M^2}{2W_{th}},~E_{\Delta}=\displaystyle\frac{W^2_{th}-m^2+M^2}{2W_{th}},~W_{th}=2m+m_{\phi},$$
where $m$, $M$ and $m_{\phi}$ are the masses of  $N$, $\Delta$ and $\phi$-meson respectively.
We used the following expression for the density matrix of the unpolarized $\Delta$-isobar:
$$\rho_{ab}^{(\Delta)}=\overline{\Delta_a\Delta_b}= \displaystyle\frac{2}{3}(\delta_{ab}-
\displaystyle\frac{i}{2}\epsilon_{abc}\sigma_c),$$
where the overline denotes the sum over the $\Delta$-polarization.

The general parametrization of the threshold matrix element, Eq. (\ref{eq:mel}), allows to calculate any polarization observable, for the considered processes. Taking into account the specificity of this reaction, which can only occur as a second step in ion-ion collisions, the most interesting (and measurable) polarization observable is the density matrix of the $\phi$-meson, which can be written (for S-state final particles) as:
$$\rho_{ab}^{(\phi)}=\hat{k_a}\hat{k_b}+
{\cal A}(\delta_{ab}-3\hat{k_a}\hat{k_b}),$$
where ${\cal A}$ is a real coefficient, which drives the angular dependence of the decay products of the $\phi$-meson. For example, in case of $\phi\to K\overline{K}$-decay, one finds $W(\theta)\simeq 1+{\cal B}\cos^2\theta$, with
\begin{equation}
{\cal B}=\displaystyle\frac{1-3{\cal A}}{{\cal A}}=-1+
\displaystyle\frac{4|f_2+f_3|^2}{|f_1-f_2|^2+3|f_1+f_2|^2},
\label{eq:eq4}
\end{equation}
where $\theta$ is the K-meson production angle (in $\phi$-rest system) with respect to the direction of the initial $\Delta$-momentum.

\section{The dynamics for the $\lowercase{t}$-channel}

\begin{figure}
\mbox{\epsfysize=15.cm\leavevmode \epsffile{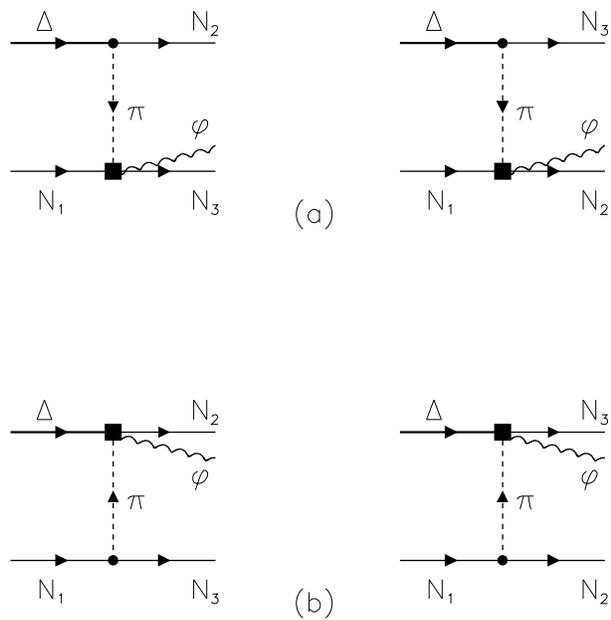}}
\vspace*{.2 truecm}
\caption{Feynman diagrams for $t-$channel $\pi$-exchanges correxponding to the subprocess $\pi+N\to N+\phi$ (a), and $\pi+\Delta\to N+\phi$ (b).
}
\label{fig:fig1}
\end{figure}

The parametrization of the spin structure of the threshold matrix elements given above, is based on fundamental symmetry properties of the strong interaction. It is therefore model independent and can be applied to any reaction mechanism. Following the analogy with the process $p+p\to p+p+\phi$, we will consider here different $t-$channel  exchanges (Fig. \ref{fig:fig1}), where only states with $I=1$ are allowed: $\pi$, $\rho$, etc.

The spin structure of the corresponding matrix elements for the subprocesses  $\pi+N(\Delta)\to N+\phi$ can be established using the conservation of the total angular momentum and $P-$parity.

\noindent\underline{The subprocess $\pi+N\to N+\phi$}

At the reaction threshold, where the $\phi$-meson is produced in the $S$-state, the following partial transitions are allowed: 
$\ell_{\pi}=0\to {\cal J}^{ P}=1/2^-$ and 
$\ell_{\pi}=2\to {\cal J}^{ P}=3/2^-$, ($\ell_{\pi}$ is the pion orbital momentum), so the matrix element can be written as follows:
\begin{equation}
{\cal M}_{th}(\pi N\to N\phi)= {\chi}^{\dagger}_2\left (h_1\vec{\sigma}\cdot\vec  U^*+h_2\vec{\sigma}\cdot\hat{\vec k }\hat{\vec k }\cdot\vec  U^*\right )\chi_1,
\label{eq:ms1}
\end{equation}
where $h_1$ and $h_2$ are the two complex partial amplitudes.
 
\noindent\underline{The subprocess $\pi+\Delta\to N+\phi$}

At the reaction threshold, this process is characterized by three possible partial transitions:  $\ell_{\pi}=0\to {\cal J}^{ P}=3/2^-$, and $\ell_{\pi}=2\to {\cal J}^{ P}=1/2^-$, and $3/2^-$, with the following matrix element:
\begin{equation}
{\cal M}_{th}(\pi \Delta\to N\phi)= 
h_1'({\chi}^{\dagger} \vec{\Delta}\cdot\vec U^*)+
h_2'({\chi}^{\dagger} \vec{\Delta}\cdot\hat{\vec k })
\hat{\vec k }\cdot\vec  U^*
+ih_3'( {\chi}^{\dagger}  \vec{\sigma}\cdot\hat{\vec k }\vec{\Delta} \cdot\hat{\vec k }\times\vec  U^*),
\label{eq:ms2}
\end{equation}
where $h_1'$, $h_2'$ and $h_3'$ are the threshold partial amplitudes for the process $\pi+\Delta\to N+\phi$, and $\chi$ is the two-component spinor of the final nucleon.

Let us consider firstly the contribution of the diagrams 1(a), which are dominated by the known (and measurable) process  $\pi+N\to N+\phi$. The corresponding matrix element (taking into account the necessary antisymmetrization with respect to the produced nucleons in $\Delta+N\to N+N+\phi$), can be written as:
$${\cal M}={\cal M}_1-{\cal M}_2$$
$$
{\cal M}_1= \displaystyle\frac{g_{\Delta N\pi}}{t-m^2_{\pi}}
({\chi}^{\dagger}_2 \vec{\Delta}\cdot\hat{\vec k })
\left [\chi^{\dagger}_3(h_1\vec{\sigma}
\cdot\vec U^*+h_2\vec{\sigma}\cdot\hat{\vec k }\hat{\vec k }\cdot
\vec  U^*) \chi_1\right ],
$$
\begin{equation}
{\cal M}_2= \displaystyle\frac{g_{\Delta N\pi}}{t-m^2_{\pi}}
({\chi}^{\dagger}_3 \vec{\Delta}\cdot\hat{\vec k })
\left [\chi^{\dagger}_2(h_1\vec{\sigma}\cdot\vec U^*+h_2\vec{\sigma}\cdot\hat{\vec k }\hat{\vec k }\cdot
\vec  U^*) \chi_1\right ],
\label{eq:ms2b}
\end{equation}
where $g_{\Delta N\pi}$ is the $\Delta N\pi$ coupling constant, for the decay 
$\Delta \to N+\pi$, which determines the corresponding width as:
$$\Gamma (\Delta \to N\pi)=
\displaystyle\frac{g_{\Delta N\pi}^2}{24\pi}\displaystyle\frac{q^3}{M^2},~q^2=E_{\pi}^2-m_{\pi}^2,~
E_{\pi}=\displaystyle\frac{M^2+m_{\pi}^2-m^2}{2M}.$$
In the threshold region for 
$\Delta+N \to N+ N+\pi$, the  pion propagators in ${\cal M}_1$ and ${\cal M}_2$ are identical. This approximation is consistent with the previous  S-wave considerations. A difference in these propagators, which appears outside the threshold region, generates P- and higher partial waves of the produced $NN\phi$-system.

Using the Fierz-transformation, in its two-component form, one can transform the matrix element ${\cal M}$, Eq. (\ref{eq:ms2b}), to the canonical ($s$-channel) parametrization, Eq. (\ref{eq:mel}). For example, the two contributions to 
${\cal M}_1$ can be written as follows:
\begin{eqnarray}
&-(\chi^{\dagger}_2\vec\Delta\cdot\hat{\vec k })~(\chi^{\dagger}_3 \vec{\sigma}
\cdot\vec A{\chi}_1)=&
\displaystyle\frac{1}{2}\left [-(\tilde{\chi}_1~\sigma_y ~\vec{\sigma}
\cdot\vec A\vec\Delta\cdot\hat{\vec k })~(\chi^{\dagger}_3
\sigma_y \tilde{\chi}^{\dagger}_2 )-\right .\nonumber \\
&& (\tilde{\chi}_1~\sigma_y\vec\Delta\cdot\hat{\vec k })~(\chi^{\dagger}_3\vec{\sigma}
\cdot\vec A \sigma_y \tilde{\chi}^{\dagger}_2 )+\label{eq:mppf} \\&&
\left . i\epsilon_{ikl}A_i(\tilde{\chi}_1~\sigma_y ~\sigma_{\ell} \vec\Delta\cdot\hat{\vec k })~(\chi^{\dagger}_3 \sigma_k\sigma_y\tilde{\chi}^{\dagger}_2 )\right ],\nonumber 
\end{eqnarray}
for any vector $\vec A$. Taking here $\vec A=\vec U^*$ or 
$\vec A=\hat{\vec k }(\hat{\vec k }\cdot\vec U^*)$, one can find the following expressions for the partial threshold amplitudes $f_1-f_3$ of the process $\Delta+N\to N+N+\phi$ , which hold in framework of the considered model:
\begin{equation}
f_1=0,~f_2=h_1\displaystyle\frac{g_{\Delta N\pi}}{t-m^2_{\pi}},
~f_3=h_2\displaystyle\frac{g_{\Delta N\pi}}{t-m^2_{\pi}}.
\label{eq:amf}
\end{equation}
From these expressions one can see that both amplitudes  $h_1$ and $h_2$ of the subprocess $\pi+N\to N+\phi$ contribute and the differential cross section for $\Delta+N\to N+N+\phi$, can be written as (in a particular normalization):
\begin{equation}
\displaystyle\frac{d\sigma}{d\omega}(\Delta N\to NN\phi)=\displaystyle\frac{N}{3}
\left ( 2|f_2|^2+|f_2+f_3|^2\right )=
\displaystyle\frac{N}{3}
\left(\displaystyle\frac{g_{\Delta N\pi}}{t-m^2_{\pi}}\right )^2
\left( 2|h_1|^2+|h_1+h_2|^2\right ).
\label{eq:eq9}
\end{equation}

Taking into account that in the considered model the parameter ${\cal B}$, Eq. (\ref{eq:eq4}), has the form:
$$1+{\cal B}= \displaystyle\frac{|f_2+f_3|^2}{|f_2|^2},$$
one can find that the simultaneous measurements of the coefficient ${\cal A}$ and the differential cross section ${d\sigma}/{d\omega}$ allow to find the module of the amplitudes:
$$|f_2|^2=N^{-1}\displaystyle\frac{3}{{\cal B}+3}\displaystyle\frac{d\sigma}{d\omega},$$
$$|f_2+f_3|^2=N^{-1}\displaystyle\frac{3({\cal B}+1)}{{\cal B}+3}\displaystyle\frac{d\sigma}{d\omega}.$$
Note that the differential cross section for the subprocess $\pi+N\to N+\phi$ has the following form, in terms of the partial amplitudes $h_1$ and $h_2$:
\begin{equation}
\displaystyle\frac{d\sigma}{d\omega}(\pi N\to N\phi)\simeq 3|h_1|^2+2 {\cal R}e~h_1h_2^*+|h_2|^2=
2|h_1|^2+|h_1+h_2|^2.
\label{eq:eq10}
\end{equation}
Therefore, in framework of the considered model of $\pi$-exchange, 
the differential cross sections for $\Delta+N\to N+N+\phi$ and $\pi+N\to N+\phi$ are proportional in the near threshold region, and the following relation holds \cite{Li76}:
\begin{equation}
\displaystyle\frac{\sigma(\Delta N\to NN\omega)}{\sigma(\Delta N\to NN\phi)}=
\displaystyle\frac{\sigma(\pi N\to N\omega)}{\sigma(\pi N\to N\phi)}.
\label{eq:prop}
\end{equation}
But it is not the case for the process $p+p\to p+p +\phi(\omega)$ or $n+p\to n+p +\phi(\omega)$, which can also be treated in framework of one-pion exchange \cite{Si00}. For the reaction $p+p\to p+p +\phi$, in the threshold region, the spin structure of the matrix element is characterized by a single contribution \cite{Re97a}, corresponding to singlet-triplet transition, of the following form:
\begin{equation}
{\cal M}_{th}(pp\to pp \phi)=f(\tilde{\chi}_2^{\dagger}\sigma_y\vec{\sigma}\cdot\hat{\vec k }\times\vec  U^*\chi_1)
(\chi^{\dagger}_4\sigma_y\tilde \chi^{\dagger}_3),
\label{eq:eq11}
\end{equation}
where $f$ is the corresponding amplitude, so that:
\begin{equation}
\displaystyle\frac{d\sigma}{d\omega}(pp\to pp \phi)\simeq 2|f|^2.
\label{eq:eq12}
\end{equation}
The $\pi$-exchange for $p+p\to p+p +\phi$ is described by four Feynman diagrams (to have the correct symmetry properties with respect to the initial and final identical protons). It is possible to prove that:
\begin{equation}
f=\displaystyle\frac{h_1}{t'-m^2_{\pi}}g_{\pi^0 pp},
\label{eq:eq13}
\end{equation} 
where $t'=-mm_{\phi}$ is the threshold value of the momentum transfer squared, for {\it elastic} $\phi$-production. One can see that for the process 
$p+p\to p+p +\phi$ only the $h_1$ amplitude contributes, therefore the proportionality of cross sections, Eq. (\ref{eq:prop}), does not hold for $pp$-collisions. 

The complete determination of the amplitudes $h_1$ and $h_2$ requires the study of polarization phenomena for the process $\pi+N\to N+\phi$, in the near threshold region. The simplest observable is the polarization of the $\phi$-meson, determined by the single parameter ${\cal B}^{(\pi)}$, which can be expressed in terms of the partial amplitudes 
 $h_1$ and  $h_2$ as:
$${\cal B}^{(\pi)} =1+\displaystyle\frac{|h_1+h_2|^2}{|h_1|^2}.$$
Therefore the measurement of the unpolarized cross section, $(d\sigma/d\Omega)_0$, and of the coefficient ${\cal B}^{(\pi)}$ allows to determine both amplitudes, through the following formulas:
$$|h_1|^2=\displaystyle\frac{1}{2(3+{\cal B^{(\pi)}})}
\left (\displaystyle\frac{d\sigma}{d\Omega}\right )_0,
~|h_1+h_2|^2=\displaystyle\frac{1+{\cal B^{(\pi)}}}{2(3+{\cal B^{(\pi)}})}
\left (\displaystyle\frac{d\sigma}{d\Omega}\right )_0.$$
 
Using Eqs. (\ref{eq:eq9}), (\ref{eq:eq10}), (\ref{eq:eq12}), and (\ref{eq:eq13}) one can find the following formula for the ratio of the total threshold cross sections of the processes $\Delta+N\to N+N+\phi$ and $p+p\to p+p+\phi$ in terms of the coupling constants and of the ratio $x= h_2/h_1$ of the two possible amplitudes for the $\pi+N\to N+\phi$ process:
\begin{equation}
{\cal R}=\displaystyle\frac{\sigma (\Delta N\to NN\phi)}{\sigma (pp\to pp \phi)}=\displaystyle\frac{1}{12}
\displaystyle\frac{g_{\Delta N\pi}^2}{g_{N N\pi}^2}\left(\displaystyle\frac{t'-m^2_{\pi}}{t-m^2_{\pi}}\right )^2 g(x),~g(x)=3+2{\cal R}e x+|x|^2,
\label{eq:eq14}
\end{equation} 
where in the integration over the final $2p$-system in $p+p\to p+p +\phi$  we have taken into account the identity of the final protons. 

The complex parameter $x$ drives also the coefficient ${\cal B}$ for the decay 
$\phi \to K^+K^-$ see Eq. (\ref{eq:eq4}): ${\cal B}=2{\cal R}e  x+|x|^2,$ so that  $3+{\cal B}=g(x)$. One  can see that these two observables are not independent, in the framework of the considered model. The comparison of the cross sections for $pp$ and $\Delta N$-collisions in
Eq (\ref{eq:eq14}) has to be done at the same excitation energy $W-W_{th}$, where $s=W^2$, and $W$  is the total invariant energy of the colliding particles, $W_{th}=2m+m_{\phi}$. 

One can show that the threshold value for the variable $t$ in the case of the reaction $\Delta+N\to N+N+\phi$ is smaller in absolute value in comparison with $p+p\to p+p+\phi$ , due to the different masses of $\Delta$ and $N$:
\begin{equation}
t=-mm_{\phi}+(M^2-m^2)\displaystyle\frac{m_{\phi}+m}{m_{\phi}+2m}.
\label{eq:tmm}
\end{equation}
As a result, the pion pole in $\Delta+N\to N+N+\phi$ is closer to the physical region, so the ratio
$$\left(\displaystyle\frac{t-m_{\pi}^2}{mm_{\phi}+m_{\pi}^2}\right )^2\simeq 4 $$
will increase the relative value of the $\Delta N$-cross section. Due to the instability of $\Delta$, this pion pole, for $\Delta+N\to N+N+\phi$ could be in the physical region, in some kinematical interval \cite{Pe96}. But, as we can see from Eq. (\ref{eq:tmm}), this can not happen in the near threshold region.

\begin{figure}
\mbox{\epsfysize=15.cm\leavevmode \epsffile{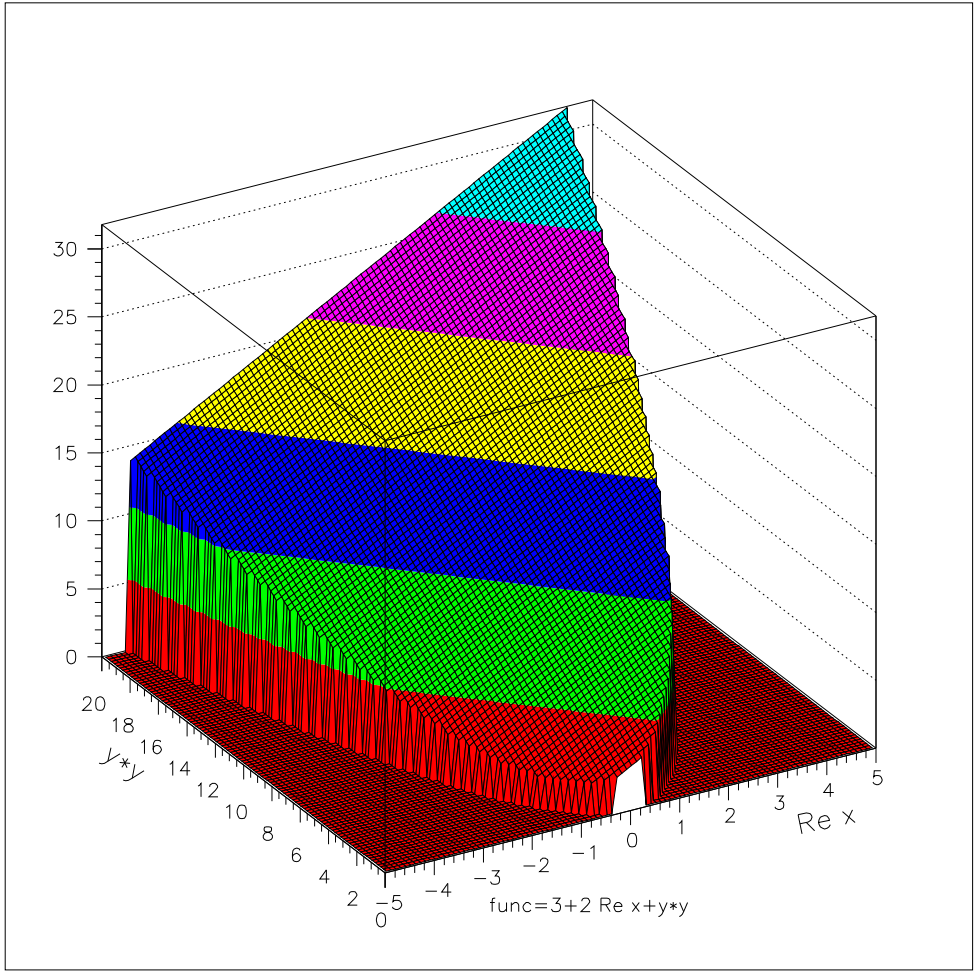}}
\vspace*{.2 truecm}
\caption{Dependence of $g(x)$ as function of  ${\cal R}e$ and $|x|$.}
\label{fig:fig2}
\end{figure}

Fig. \ref{fig:fig2} shows the $x$-dependence of $g(x)$ as a function of the two independent kinematical variables,  ${\cal R}e$ and $|x|$, with the evident conditions $|{\cal R}e x|\le |x|$, $|x|\ge 0$, ${\cal R}e x$ being positive or negative.

Taking $\Gamma(\Delta\to N\pi)\simeq$ 120 MeV and $g^2_{\pi NN}/4\pi=15$ one 
 can find  ${\cal R}\ge 1$, i.e. the processes $p+p\to p+p+\phi$ and $\Delta+N\to N+N+\phi$  have comparable cross section, in the threshold region. A more accurate estimation for ${d\sigma}/{d\omega}$ and ${\cal R}$ could be done, knowing the threshold partial amplitudes $ h_1$ and $h_2$. In literature \cite{Si97} it is possible to find an estimation of the energy dependence of the total cross section (with unpolarized particles) in the framework of a very specific resonance model where a single nucleon resonance contributes. The mass and width of this 'effective' resonance were determined by fitting the available experimental data \cite{Co70}. Such model, however, can not help to determine the partial amplitudes $ h_1$ and $h_2$ without additional assumptions, concerning the spin and parity of the resonance $R$ and its coupling constants for the vertex $R\to N+\phi$.

In principle it is possible to develop a more complicated model for the $\pi+N\to N+\phi$ process, which takes into account different contributions: vector meson exchange in $t$-channel, $N$ and $N^*$ contributions in $s$- and $u$-channels \cite{Ti00}. But even near the reaction threshold several resonances with ${\cal J}^P=1/2^-$ and $3/2^-$contribute, introducing unknown parameters, coupling constants and form factors. Only in the framework of a quark model \cite{Ti01} definite predictions can be done.

\begin{figure}
\mbox{\epsfysize=15.cm\leavevmode \epsffile{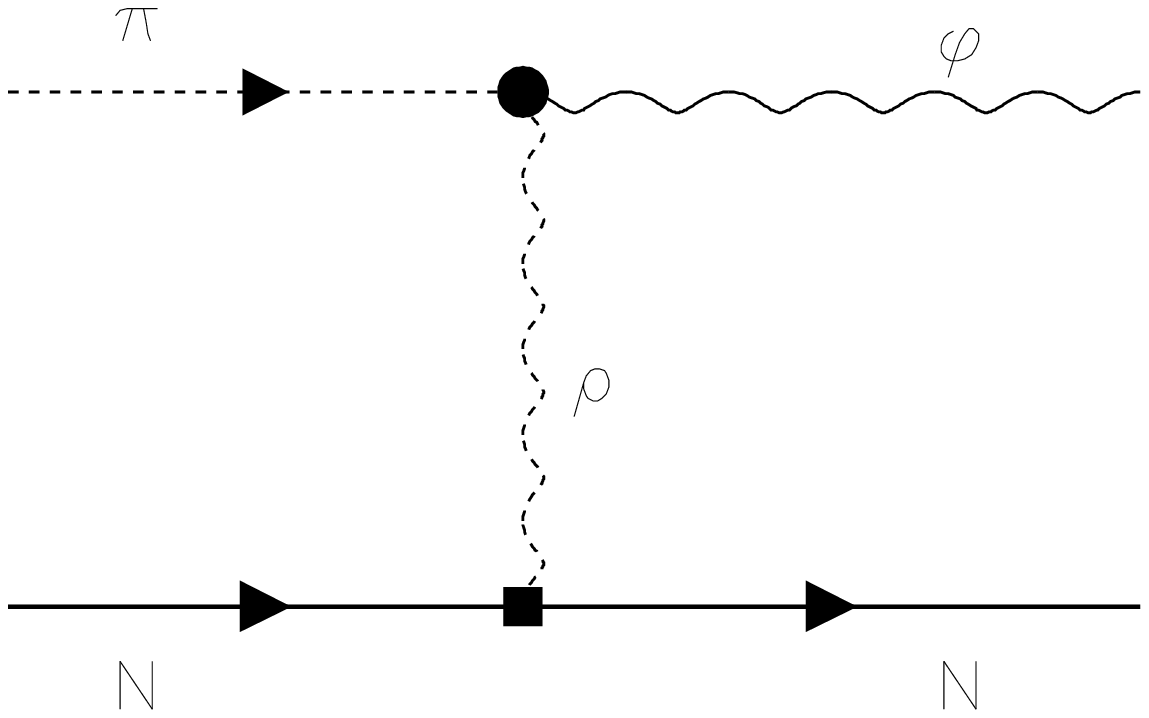}}
\vspace*{.2 truecm}
\caption{Feynman diagram for $t$-channel $\rho$-exchange 
for the process $\pi+N\to N+\phi$.}
\label{fig:fig3}
\end{figure}

Let us apply the $\rho$-exchange for the 'elementary' process $\pi+N(\Delta)\to N+\phi$, Fig. \ref{fig:fig3}, to estimate the value of the ratio $x$. One can easily see that, in the near threshold region, such model gives $x=-1$ and the minimal value for the ratio of cross section ${\cal R}$. In this case one finds ${\cal A}=-1$, with a $\sin^2\theta$-angular dependence of the decay products in $\phi\to K+\overline{K}$, which is an evident property of the $\pi\rho\phi$-vertex. This last result does not depend on the different ingredients of the reaction mechanism, such as the coupling constants and the phenomenological form factors in the hadronic vertexes, but it depends on the mechanism chosen for  the process $\pi+\Delta\to N+\phi$. However, as in the case of $\pi+N\to N+\phi$, one can expect that these contributions (nucleon exchanges in $s$-and $u$-channels, nucleon resonances $N^*$-excitations) are small in the near threshold region.

It is possible to analyze in the same way the $\rho$-exchange for the process $\Delta+N\to N+N+\phi$. The corresponding analytical analysis can be done in terms of the partial threshold amplitudes for the subprocesses $\rho^*+N\to N+\phi$ and
$\rho^*+\Delta \to N+\phi$, where $\rho^*$ is the virtual $\rho$-meson, with space-like four-momentum. For numerical estimation an adequate dynamical model has to be built for such exotic subprocesses. For $\rho^*+N\to N+\phi$ one can apply VDM, which connects this process to $\phi$-electroproduction in threshold conditions. The process $\rho^*+\Delta\to N+\phi$, even at threshold, is characterized by a complicated spin structure, with five independent partial amplitudes.

\section{Conclusions}
We have studied the reactions $\Delta+N\to N+N+\phi$ in the near threshold region on the basis of the most general symmetry properties of the strong interaction 
(P-parity and angular momentum conservation, as selection rules) and treating the dynamical aspects in the framework of one-boson mechanism. Let us summarize the main results of our analysis.
\begin{itemize}
\item We established the spin structure of the threshold matrix element  for the process $\Delta+N\to N+N+\phi$ in terms of three partial independent complex amplitudes. In the general case of non-coplanar kinematics, this process is described by 96 independent amplitudes, 48 in the coplanar case and 11 amplitudes for collinear kinematics.

\item One from these threshold amplitudes vanishes for the pion $t$-channel exchange - with the subprocess $\pi+N\to N+\phi$.
Therefore it is possible to predict the polarization properties of the $\phi$-mesons, produced in the reaction  $\Delta+N\to N+N+\phi$, and the ratio of the total cross section for the processes $\Delta+N\to N+N+\phi$ and 
$p+p\to p+p+\phi$ in terms of a single parameter $x$, which characterizes the relative role of production of $\phi$-mesons with transversal and longitudinal polarizations in  the process  $\pi+N\to N+\phi$.
\item The value of the parameter $x$ depends on the model chosen to describe the process $\pi+N\to N+\phi$ in the near threshold region. The $\rho$-exchange mechanism gives $x=-1$.
\item The threshold cross sections for the process $\Delta+N\to N+N+\phi$ and $p+p\to p+p+\phi$ are comparable.
\item The polarization properties of the $\phi$-meson, produced in the processes $\Delta+N\to N+N+\phi$ and 
$N+N\to N+N+\phi$ are generally different. In the last case, near threshold, the $\phi$-mesons are only transversally polarized.
\end{itemize}

These predictions are independent on the value of the $NN\phi$-coupling constants and the 
phenomenological hadronic form factors for the different meson-nucleon vertexes.

\begin{references}


\bibitem{Ba01}
F.~Balestra {\it et al.}  [DISTO Collaboration],
Phys.\ Rev.\ C {\bf 63}, 024004 (2001); Phys.\ Rev.\ Lett. {\bf 81}, 4572 (1997); 
Phys.\ Lett.\ B {\bf 468}, 7 (1999).


\bibitem{Si97}
A.~Sibirtsev, W.~Cassing and U.~Mosel,
Z.\ Phys.\ A {\bf 358}, 357 (1997).


\bibitem{Si00}
A.~Sibirtsev and W.~Cassing,
Eur.\ Phys. Journ.\ A {\bf 7}, 407 (2000).

\bibitem{Na98}
K.~Nakayama, A.~Szczurek, C.~Hanhart, J.~Haidenbauer and J.~Speth,
Phys.\ Rev.\ C {\bf 57}, 1580 (1998).

\bibitem{Na99}
K.~Nakayama, J.~W.~Durso, J.~Haidenbauer, C.~Hanhart and J.~Speth,
Phys.\ Rev.\ C {\bf 60}, 055209 (1999).

\bibitem{Ti00}
A.~I.~Titov, B.~Kampfer and B.~L.~Reznik,
Eur.\ Phys.\ J.\ A {\bf 7}, 543 (2000).

\bibitem{Fa01}A.~Faessler, V.~I.~Kukulin, I.~T.~Obukhovsky, and V.~N.~Pomerantsev,
J.Phys. G27, 1851 (2001). 

\bibitem{Ta00}
B.~Tatischeff {\it et al.},
Phys.\ Rev.\ C {\bf 62}, 054001 (2000).
\bibitem{El95} J. Ellis, M. Karliner, D. E. Kharzeev, and M. G. Sapozhnikov,
Phys. Lett. B {\bf353} (1995) 319; Nucl. Phys. A {\bf 673}, 256 (2000). 


\bibitem{OZI} S.Okubo, Phys. Lett. B {\bf 5}, 165 (1963);\\
G. Zweig, CERN Reports TH401 and TH412 (1964);\\
J. Iizuka, K. Okada, and O. Shito, Prog. Theor. Phys. Suppl. {\bf 37}, 38 (1966).  

\bibitem{Sh85}
A.~Shor,
Phys.\ Rev.\ Lett.\  {\bf 54}, 1122 (1985).

\bibitem{Ba91}
C.~Baglin {\it et al.}  [NA38 Collaboration],
Phys.\ Lett.\ B {\bf 272} (1991) 449;\\
M.~C.~Abreu {\it et al.}  [NA38 Collaboration],
Phys.\ Lett.\ B {\bf 368} (1996) 230;
Phys.\ Lett.\ B {\bf 368} (1996) 239.

\bibitem{Me93}
V.~Metag,
Prog.\ Part.\ Nucl.\ Phys.\  {\bf 30}, 75 (1993).

\bibitem{Pe96}
W.~Peters, U.~Mosel and A.~Engel,
Z.\ Phys.\ A {\bf 353}, 333 (1996).


\bibitem{Re97a}
M.~P.~Rekalo, J.~Arvieux and E.~Tomasi-Gustafsson,
Z.\ Phys.\ A {\bf 357}, 133 (1997).



\bibitem{Bo68}
L.~Bodini {\it et al.}, Nuovo Cim. A {\bf 58}, 475 (1968).

\bibitem{Co67}
A.~P.~Colleraine and U. Nauenberg, Phys.\ Rev.\ {\bf 161}, 1387 (1967).
\bibitem{Ca68} 
C. Caso {\it et al.}, Nuovo Cim. A {\bf 55}, 66 (1968).

\bibitem{Co70} 
E. Colton and E. Gellert, Phys.\ Rev.\ D {\bf 1}, 1979 (1970).

\bibitem{Ye70} 
G.~Yekutieli {\it et al.},
Nucl.\ Phys.\ B {\bf 18}, 301 (1970).

\bibitem{Al68}
S.~P.~Almeida {\it et al.},
Phys.\ Rev.\  {\bf 174}, 1638 (1968).

\bibitem{Hi99}
F.~Hibou {\it et al.},
Phys.\ Rev.\ Lett.\  {\bf 83}, 492 (1999).
\bibitem{Ab01}
S.~Abd El-Samad {\it et al.}  [COSY-TOF Collaboration],
Phys.\ Lett.\ B {\bf 522}, 16 (2001).


\bibitem{Re97b}
M.~P.~Rekalo, J.~Arvieux and E.~Tomasi-Gustafsson,
Phys.\ Rev. C {\bf 56}, 2238 (1997).


\bibitem{Ba98}
J.~T.~Balewski {\it et al.},
Phys.\ Lett.\ B {\bf 420}, 211 (1998).
\bibitem{Bi98}
R.~Bilger {\it et al.},
Phys.\ Lett.\ B {\bf 420}, 217 (1998).

\bibitem{Se98}
S.~Sewerin {\it et al.},
Phys.\ Rev.\ Lett.\  {\bf 83}, 682 (1999).

\bibitem{Ba99}
F.~Balestra {\it et al.}  [DISTO Collaboration],
Phys.\ Rev.\ Lett.\  {\bf 83}, 1534 (1999).


\bibitem{Li76}
H.~J.~Lipkin,
Phys.\ Lett.\ B {\bf 60}, 371 (1976).
\bibitem{Ti01}
A.~I.~Titov, B.~Kampfer and B.~L.~Reznik,
arXiv:nucl-th/0102032.

\end{references}
\end{document}